# Strong electron-phonon coupling in 3D tungsten nitride and coexistence of intrinsic superconductivity and topological nodal line in its 2D limit

Jianyong Chen[1*], Jiacheng Gao[2]

[1]College of science, Guilin University of Aerospace Technology, Guilin 541004, China

[2]Beijing National Laboratory for Condensed Matter Physics, and Institute of Physics, Chinese Academy of Sciences, Beijing 100190, China

Three-component fermion beyond the conventional Dirac–Weyl–Majorana classification attracts extensive attentions recently and many efforts have been paid to explore their superconductivity. Based on first-principles calculations, we systematically investigate the electron-phonon coupling (EPC) in the three-component fermion materials WN, WC and TaN. The EPC in pristine and pressured WC and TaN are to small to induce superconductivity. Electron doping can efficiently enhance the EPC strength and the predicted $T_c$s reach the value of experiments. Upon 0.6 electron/unitcell doping, the EPC strength of TaN is boosted by two orders of magnitude and $T_c$ can even be as high as 27 K, revealing the crucial role of charge doping in the formation of superconductivity observed in WC and TaN. In stark contrast, pristine WN exhibits overwhelmingly strong EPC and can be a good superconductor with a high transition temperature $T_c$ of 31 K. The strong EPC in WN are dictated by a synergistic effect of strong Fermi nesting and large deformation potential. Going down from three-dimension (3D) to three-dimension (2D), WN thin film (i.e. monolayer $W_3N_4$) is also an intrinsic superconductor with $T_c$ of 11 K. Most importantly, monolayer $W_3N_4$ host Dirac nodal lines protected by mirror symmetry in the absence of spin-orbit coupling (SOC), Including SOC, the Dirac nodal lines split into three pairs of spinful Weyl rings. These nodal lines lies closely near the Fermi level, they are pure and clean without other nontrivial bands, which is scarce in real materials and making the exotic topological properties easily accessible in experiment. The coexistence of superconductivity with high transition temperature and topological states in WN and its 2D film provide a promising platform for exploring topological superconductivity.

## I. INTRODUCTION

Topological superconductors (TSC) which are topologically distinct from conventional Bose–Einstein condensates of Cooper pairs have become a focus of interest due to their potential applications in the fault-tolerant topological quantum computation [1]. Two possible ways to access TSC have been demonstrated, one is to build artificial TSC using hybrid structures. For example, utilizing the proximity effect between an *s*-wave superconductor and a spin-nondegenerate metal [2,3]. The other promising scheme is to inducing superconductivity in nontrivial topological materials by doping or pressurizing topological insulators [4-6], Dirac semimetals [7,8], Weyl semimetals [9,10] or nodal lines semimetals [11]. Going beyond Weyl, Dirac and nodal line semimetals, there are "new fermions" characterized by three-, six-, or eightfold degenerate points protected by nonsymmorphic space group symmetries [12]. Recently, three-component fermions (the three-component fermion coexists with nodal lines in WN and WC) are proposed in the tungsten carbide (WC)-type structure, including WN, WC, TaN, MoP,

ZrTe, MoN et al [13,14] protected by the combination of rotation and mirror symmetries. Experimentally, three-component fermions are first detected in MoP [15], then the non-trivial topological surface states is observed in WC [16]. At ambient pressure, all above stoichiometric compounds are not superconductors. To explore their superconductivity, much efforts and attentions have been paid. *Z. Chi et al* find that pressure can induce superconductivity in bulk MoP with $T_c$ from 2 K to 4 K with the pressure increasing from 30 GPa to 95 GPa [17]. Similarly, TaN also turns to be a superconductor when pressure is larger than 25 GPa, $T_c$ increases monotonically with the increasing of pressure accompanying with a increasing of carrier density [18]. The maximum Tc is 4.68 K at pressure 56.3 GPa. Local superconductivity can also be induced at the point contact between a normal metal (non-magnetic Au, Pt, Ir, W, Ta and magnetic Fe, Co, Ni) tip and WC. The critical temperatures are in the range of 4~12 K and not sensitive to the metal types [19]. *W. Zhu et al* discover non-local interfacial superconductivity on the interface between WC and metallic thin films (Au, Pt, Fe, Co, Ni) [20]. Due to the superhardness of WC, the tip pressure or confinement effect can be excluded as confirmed by soft point contact testing [20]. Finally, coupling between non-superconducting metals and WC should be a key ingredient for the induced superconductivity. At the present time, the microscopic nature of the interfaces and the mechanism of superconductivity remains elusive. Worth noting is that the local superconductivity induced by tip contact and non-local superconductivity induced by metal deposition in WC are not sensitive to the magnetism of metals, or the work funtion of different metals. These strongly suggest that the superconductivity in WC is associated with their common effect, i.e. charge doping.

Considering that the tip induced and metal deposition induced superconductivity are confined at metal/WC interfaces, it is necessary to explore the properties of their thin films as compared with bulk counterparts. 3D-to-2D crossover are predicted to induce quantum phase transition depending on surface orientation, terminations and thickness of the films. Specifically, $Cd_3As_2$, a typical Dirac semimetals in its bulk form, becomes trivial insulator below seven monolayers [21] and a dual 2D topological insulator and topological crystalline insulator in the monolayer limit [22]. Another Dirac semimetals $Na_3Bi$, is experimentally found to be topological insulator for monolayer and double layer $Na_3Bi$ films on Si(111) substrate [23]. Moreover, thin film open new possibilities to tune the topological state with thickness, electric field, and proximity coupling to superconductors, ferromagnets and ferroelectric. In the bulk form, there are nodal lines in WN and WC at $k_z=0$ plane, but mixing with nontrivial states. The contribution from these trivial states significantly complicates the analysis of topological surface states and submerge their novel transport behavior. For instance, as a typical nodal lines candidate, the nodal line electronic structure indeed has been visualized with ARPES [24]. However, no exciting transport properties have been reported. Recently, high-quality nanometer-thin (10~30 nm) WC crystals with sizes larger than 10 μm are successfully synthesized by chemical vapor deposition (CVD) technique [25,26], although it is not a layered material. This success stimulate us to explore the evolution of electronic structures, topology and EPC when the dimensionality is reduced to 2D.

In this work, we calculate and analyze systematically the EPC and superconductivity of bulk WN, WC and TaN. While pristine and pressured WC and TaN show negligibly small EPC, electron doping can significantly enhance the EPC and the predicted $T_c$s are the same order with experiments. In contrast, pristine WN host extremely strong EPC and a high $T_c$ of 31 K. When WN is thinned down to 2D monolayer $W_3N_4$, unexpected topological nodal lines emerge, they are pure and clean and robust against SOC. The paper is organized as follows: Sec.II describes the details of *ab initio* calculations, Sec. III presents the results of the calculations and main analysis, and Sec. V summarizes the main conclusions of this work.

## II. COMPUTATIONAL DETAILS

First-principles calculations are performed with Quantum-ESPRESSO package [27], using Norm-conserving potentials and Perdew-Burke-Ernzerh functional [28]. Kinetic cutoff for the plane waves is 80 Ry. Brillouin zone are sampled with $16 \times 16 \times 16$ ($16 \times 16 \times 1$) grid for the electronic properties of bulk (monolayer). The dynamical matrices and the EPC are calculated using density functional perturbation theory (DFPT) [29] in the linear response regime on a $8 \times 8 \times 8$ ($8 \times 8 \times 1$) Monkhorst–Pack (MP) [30] *q*-point grids for 3D bulk and 2D monolayers. The EPC converges when adopting a fine grid of $32 \times 32 \times 1$ *k*-point grids with a Marzari-Vanderbilt smearing [31] of 0.02 Ry. EPW code [32,33] are used to obtain the anisotropical superconducting gap. The Maximally localized Wannier functions (MLWF) [34] are constructed on a uniform unshifted $16 \times 16 \times 16$ *k*-point grid for 3D bulk. N $sp^3$ and W *d* orbitals are chosen as projectors. The Matsubara frequency cutoff is set to seven times the largest phonon frequency and the Dirac delta functions are replaced by Lorentzians of widths 25 and 0.05 meV for electrons and phonons respectively. Electrons within ±200 meV from the Fermi energy are taken in to the EPC process.

$T_c$ can be determined by McMillian-Allen-Dynes formula [35,36]:

$$T_c = \frac{\omega_{\log}}{1.2} \exp\left[\frac{-1.04(1+\lambda)}{\lambda - \mu^*(1+0.62\lambda)}\right] \quad (1)$$

where $\omega_{\log}$ is logarithmic average of the phonon frequencies. The retarded Coulomb pseudopotential $\mu^*$ in equation (1) measures the strength of electron-electron interaction. $\mu^*=0.10$ is used for bulk tungsten, bulk WN and monolayer $W_3N_4$, because this value more accurately fit the experimental data. The λ and $\omega_{log}$ can be calculated through isotropical Eliashberg function [37]:

$$\lambda = 2\int_0^\infty d\omega \alpha^2 F(\omega)/\omega = \sum_{\mathbf{q}\nu} \lambda_{\mathbf{q}\nu} \quad (2)$$

$$\omega_{\log} = \exp\left[\frac{2}{\lambda}\int_0^{\omega_{\max}} \alpha^2 F(\omega)\frac{\ln(\omega)}{\omega}d\omega\right] \quad (3)$$

$$\alpha^2 F(\omega) = \frac{1}{2\pi N(E_F)} \sum_{\mathbf{q}\nu} \frac{\gamma_{\mathbf{q}\nu}}{\omega_{\mathbf{q}\nu}} \delta(\omega - \omega_{\mathbf{q}\nu}) \quad (4)$$

where $\alpha^2 F(\omega)$ is Eliashberg function and $N(E_F)$ is the DOS at the Fermi energy, cumulative EPC is given by

$$\lambda(\omega) = \int_0^\omega d\omega' \alpha^2 F(\omega') \quad (5)$$

The phonon linewidth is defined as:

$$\gamma_{\mathbf{q}\nu} = 2\pi\omega_{\mathbf{q}\nu} \sum_{\mathbf{k}jj'} \left|g_{\mathbf{k}+\mathbf{q}j',\mathbf{k}j}^{\mathbf{q}\nu}\right|^2 \delta(\varepsilon_{\mathbf{k}j} - \varepsilon_F)\delta(\varepsilon_{\mathbf{k}+\mathbf{q}j'} - \varepsilon_F) \quad (6)$$

where the double Dirac delta functions restricts the sum of *e*-ph matrix elements to electronic states at the Fermi level, $\varepsilon_{kj}$ is the one-electron band energy with momentum $k$ and band index $j$, screened electron-phonon matrix element is defined as:

$$g_{k+qj,kj}^{q\upsilon,jj} = \sum_{R,\upsilon} \frac{\eta_{q\upsilon}(R,\upsilon)}{\sqrt{2M_R\omega_{q\upsilon}}} \langle k+q,j' \left| \frac{\delta V_{eff}}{\delta R_{\upsilon}} \right| k,j \rangle \qquad (7)$$

which describes the electron-phonon matrix element for the scattering between the electronic states $|k,j\rangle$ and $|k+q,j'\rangle$ through the phonon mode $|q,\upsilon\rangle$. $\delta V_{eff}/\delta R_{\upsilon}$ denotes the change in the total effective crystal potential with respect to the displacement of atom $\nu$. The relationship between $\lambda_{q\nu}$ for mode ν at vevector **q** and phonon linewidth are $\lambda_{q\nu} = \gamma_{q\nu}/(\pi h N(E_F)\omega_{q\nu}^2)$.

## III. RESULTS AND DISCUSSIONS

### A. Electron-phonon coupling and superconductivity in 3D WN, WC and TaN

WN, WC and TaN all crystallize in a noncentrosymmetric hexagonal lattice with space group $P\bar{6}m2$ (no. 187). The metal atoms and non-metal atoms are at the 1$d$ (1/3,2/3,1/2) and 1$a$ (0,0,0) Wyckoff position respectively. The theoretically optimized lattice constants are in good accordance with experiments (given in the parentheses), $a = b = 2.806$ (2.890) Å, $c = 2.841$ (2.830) Å for WN [38], $a = b = 2.891$ (2.907) Å, $c = 2.820$ (2.837) Å for WC [39], $a = b = 2.810$ (2.930) Å, $c = 2.785$ (2.880) Å for TaN [40]. The crystal structures possess the rotational symmetry $C_{3z}$ and the mirror symmetries $M_y$ and $M_z$, which are necessary to protect the topological fermions. Fig. 1(a, f, k) present the phonon dispersion decorated with red dots proportional to the partial EPC strength $\lambda_{q\upsilon}$, isotropic Eliashberg spectral functions $\alpha^2F(\omega)$ and cumulative frequency-dependent $\lambda(\omega)$ of WC, WN and TaN. The phonon spectra agree well with previous calculations [41,42]. The phonon spectra of WN softens as compared with WC and TaN, besides, there are Kohn anomaly for optical modes at K and M points. We recalculate the phonon dispersion of WN by replacing the mass of N with C as shown with black dashed lines in Fig. 1(a), the optical modes are still lower than that of WC, this testing exclude the mass induced lowering of frequency. The remarkable difference between WN and WC (TaN) is the overwhelming large total λ of 1.37 as compared with 0.14 and 0.02 for WC and TaN (see Fig. 1(a, f, k) and Tab. 1). The dominate contribution to EPC are denoted as α, β, δ, γ and ε. By comparing the phonon dispersion of bulk WN with WC and TaN, one can clearly figure out that these phonons show obvious softening. The vibrational pattern in real space of α and β modes are shown in Fig. 2(b,c) and others are shown in Fig. S1 in Supplemental Material. Worth mention is that all the six modes contribute to EPC, this scenario is different from MgB$_2$ in which a single mode dominates λ [43]. α, β and ε phonons which contribute 70% of the total λ are dominated by (almost purely) vibration of W atoms.

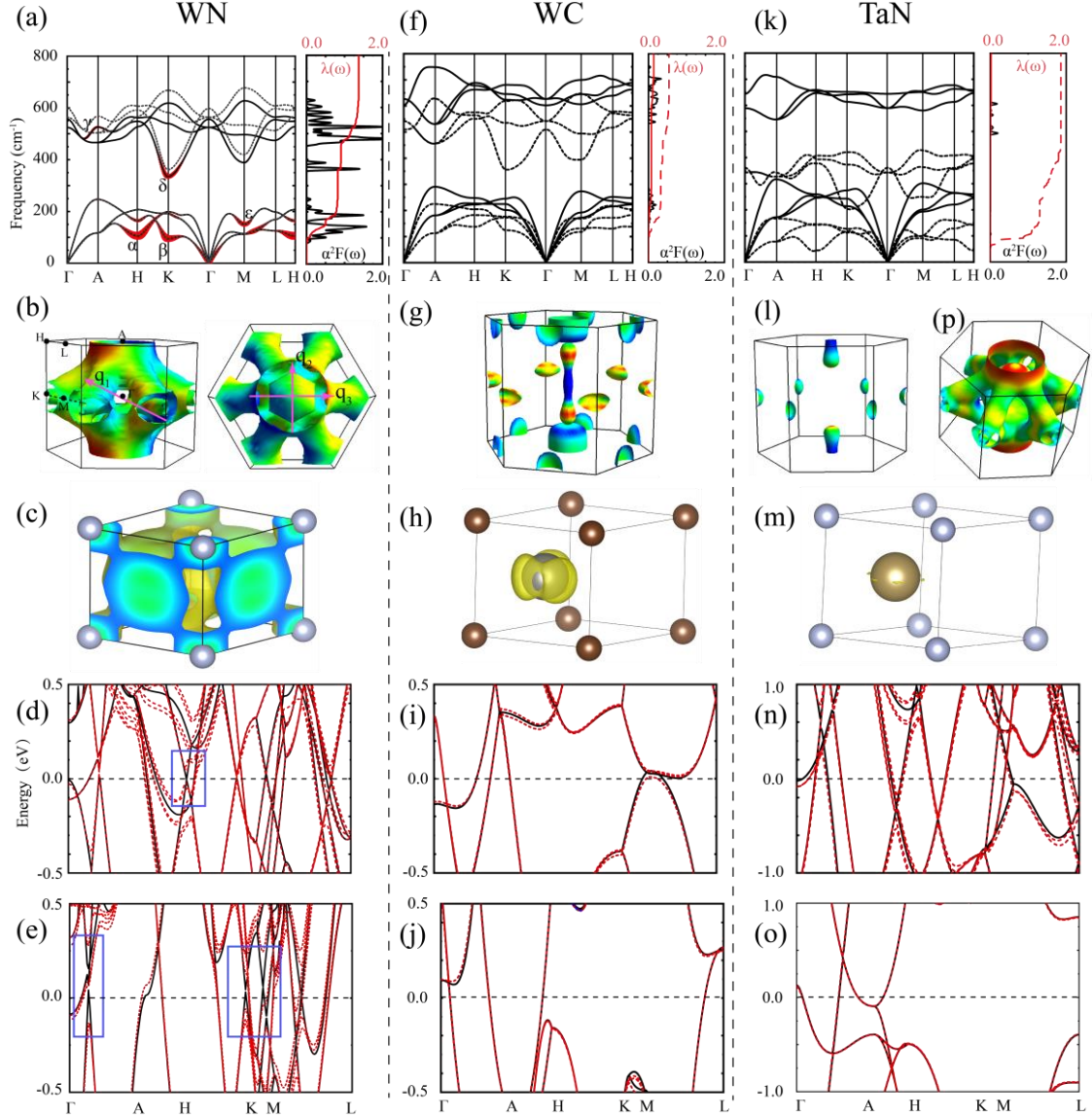

FIG. 1. (a) Phonon dispersion (solid black lines) of pristine WN decorated with red dots proportional to the partial EPC strength $\lambda_{\mathbf{q}\nu}$ (left panel), the dashed black lines are results obtained by replacing the mass of N with C. Isotropic Eliashberg spectral functions $\alpha^2F(\omega)$ (black solid lines) and cumulative frequency-dependent $\lambda(\omega)$ (red dashed line) (right panel). (b) Fermi surface and the major nesting wave vectors (purple arrows) connecting the strong nesting pieces of Fermi surface. $q_1$ represents nesting vector of $\Gamma$-$H$, $q_2$, $q_3$ represent nesting vector of $\Gamma$-$K$ and $\Gamma$-$M$. (c) Partial charges of states at Fermi level, the isosurface level is set as 0.00013 electron/Bohr$^3$. (d) Calculated band structures of bulk WN in a $3 \times 3 \times 2$ supercell before (solid lines) and after (dashed lines) being distorted by the $\alpha$ phonon at $H$ point of the primitive cell. (e) Calculated band structures of WN in a $3 \times 3 \times 1$ supercell before (solid lines) and after (dashed lines) being distorted by the $\beta$ phonon at $K$ point of the primitive cell. The displacements of atom in (b) and (c) are all 0.06 Å. The same quantities for (f-j) pristine bulk WC and (k-o) TaN. The dashed lines in (f) are results for 1.0 electron/unitcell doped WC. The dashed lines in (k) are results for 0.6 electron/unitcell doped TaN. The zero of energy in (d-e, i-j, n-o) are set at the Fermi energy of the unperturbed system.

To understand the tremendous difference between WN and WC/TaN, we first display their Fermi surfaces in Fig. 1 (b,g,l). There are large pieces of Fermi surface of WN that nesting with each other, $q_1$ represents nesting vector of $\Gamma$-$H$, $q_2$, $q_3$ represent nesting vector of $\Gamma$-$K$ and $\Gamma$-$M$. Those nesting vectors coincide with the positions of soft phonons. While for WC and TaN, the Fermi surfaces are small and weakly nest. According to equation (6) and (7), strong Fermi nesting is a prerequisite for strong EPC, another essential requirement is large screened electron-phonon matrix (or deformation potential). As a typical case, remarkable EPC can only be obtained in doped graphene by exerting notable strain so as to increase sharply the deformation potential [44]. Partial charge of states at Fermi level in real space provides a direct visual judgement of whether deformation potential are weak or strong. For bulk WC and TaN, partial charges near Fermi level are located closely around W (or Ta) atoms, there is no overlap of this states with nearest W/Ta or C/N atoms (Fig. 1(h,m)), hence the displacements induced by any phonon in these compound can not change the overlap of the states near Fermi level, which are confirmed by phonon distorted electronic structures in Fig. 1(i,j,n,o). This naturally results in negligibly small EPC strength. For bulk WN, the partial charges near Fermi level are quite delocalized, they fulfill the void space between W and its nearest neighbor and the space between W and N atoms (Fig. 1(c)). This means that atom displacement induced by any of the phonons will change the overlap of these states, which promises a large EPC strength. Especially, the overlap between W-W atoms are larger that that between W and N atoms, this explains why the four modes with large partial $\lambda_{\mathbf{q}\upsilon}$ are associated mainly with W vibrations. Furthermore, the binding are more compact in *xy* plane (see Fig. 1(c)), so vibrations along *xy* plane contribute more EPC than those along *z* axis (see vibration patter of α and ε).

To explicitly display the strong deformation potential in WN, Fig. 1(d,i,n) and Fig. 1(e,j,o) show the calculated band structures of the three compound before and after being distorted by α and β modes respectively. For bulk WN, under the distortion of α phonon, the band degeneracy at the *H* point of the supercell is removed obviously. Similarly, under the distortion of β phonon, the band degeneracy along $\Gamma$-$A$, $K$-$H$ and $K$-$M$ of the supercell are removed. The splitting energy of above bands are 61, 194, 193 and 285 meV respectively, one of which is even larger than the splitting energy of 180 meV induced by the $E_{2g}$ phonon in $MgB_2$ [45]. The phonon induced strong breaking of electronic degeneracies plays a dominant role in the phonon softening and leads to anomalous large EPC strength [46,47]. In stark contrast, the electronic structures of WC and TaN supercell are almost unaffected by the perturbation of α and β phonon as shown in Fig. 2(i,j) and Fig. 2(n,o).

The $N(E_F)$=5.56 states/spin/Ry/unitcell of bulk MoP is comparable with $N(E_F)$=6.24 states/spin/Ry/unitcell of bulk WN and its Fermisurface also display strong nesting along $\Gamma$-$K$ and $\Gamma$-$M$ [17]. However, our previous calculation show that its total λ is only 0.28 [48]. We can understand this difference by looking their spatial charge density at Fermi level. Under the same isosurface level, as 4*d* orbitals in Mo are more localized than the 5*d* orbitals in W, the charges in in MoP locate closely around Mo atoms [48], results in nearly zero deformation potential. As the electronic states at Fermi level are dominated by W *d* states and the vibration modes are mostly originate from W atoms, moreover, the distance of the nearest neighbor in tungsten is almost the same with that in bulk W, one may expect that the EPC strength of WN should be comparable with that of bulk W, whose superconducting transition temperature is only 0.015K [49]. (Our calculations show that the total λ and $T_c$ for bulk W are 0.27 and 0.04 K respectively (see Tab. 1), agrees well with experiments.) In fact, The N($E_F$) of bulk WN (6.24 states/spin/Ry/unitcell) is more than two times the value of tungsten (2.96 states/spin/Ry/cell). Most importantly, the Fermi nesting in bulk WN are much stronger than that in tungsten [50], this strong nesting allows more electron and phonons participating the coupling process, bringing pronounced softening of phonons, both of these are beneficial for enhancing EPC.

Next, we strive to get some clues of the observed superconductivity in WC [20] and TaN [18] by calculating the EPC of WC and TaN under hydrostatic/unaixal pressures or electron dopings. By applying pressure, the total λ are too small to induce superconductivity regardless of large hydrostatic pressure of 35 GPa or uniaxial pressure as large as 50 GPa (see Tab. 1). The λ of WC film ($W_2C$, $W_4C_3$ and $W_5C_6$ ) are also too small to induce superconductivity (Supplemental Material Tab. S1), implying that interface interaction induced structural changes are not enough to induce superconductivity. Fig. S2-3 in Supplemental Material present the partial charge at 2 eV and 4 eV above the Fermi level within the rigid band approximation, as those charges are more delocalized, the EPC may be enhanced if heavy electron doping can be realized in WC and WN via chemical or gating strategy. With 1.0 and 0.2 electron/unitcell doping for WC and TaN, the phonon dispersion presents obvious softening (see Fig. 1 (f,k)), the total λ of WC (TaN) are enhanced to 0.54 (0.41) and the estimated $T_c$ reaches 5.57 (1.48) K, which is in the same order of experimental values (see Tab. 1). With 0.6 electron/unitcell doping, N($E_F$) in TaN increases to 14 times that of the pristine one, Fermi surface topology (see Fig. 1(p)) are expectedly the same with pristine WN, leading to remarkable phonon softening (the optical modes downshift by even 400 $cm^{-1}$), especially near *H*, *K* and *M* points. The predicted $T_c$ in TaN can even reach 26.98 K. These results reflect that doping play a vital role in inducing superconductivity in WC and TaN, which explain the close relationship between $T_c$ and carrier density found in experiments [18] [20].

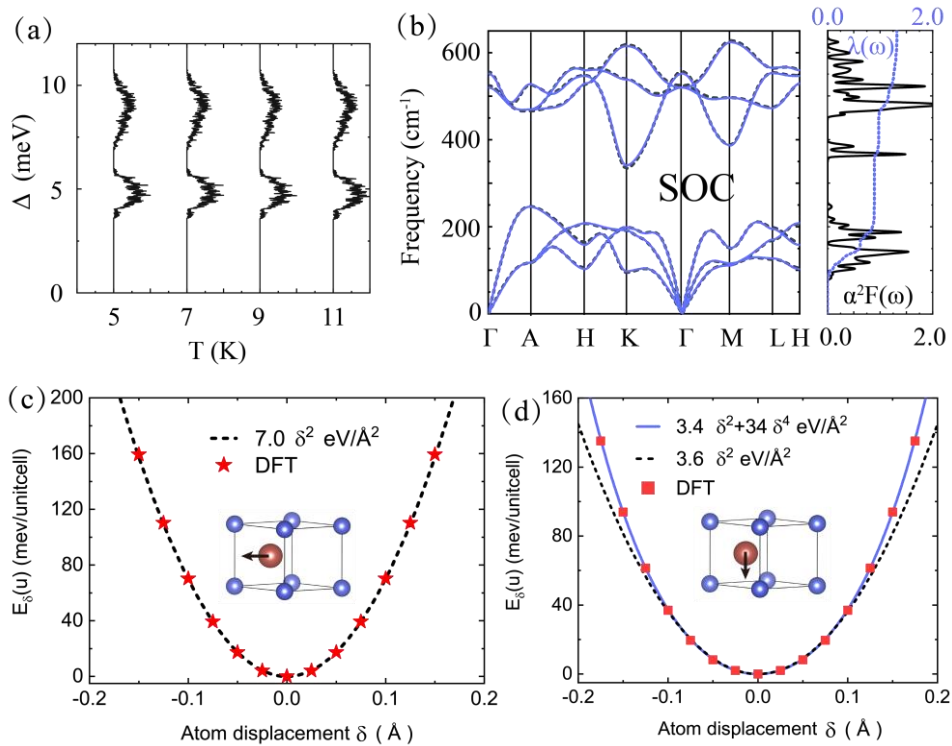

FIG. 2. (a) Energy distribution of the anisotropic superconducting gap Δ as a function of temperature from 5 K to 11 K. (b) Phonon dispersion (solid blue lines), isotropic Eliashberg spectral functions $\alpha^2 F(\omega)$ (black solid lines) and cumulative frequency-dependent $\lambda(\omega)$ (blue dashed line) calculated with SOC included, the phonon spectra calculated without SOC (dashed black lines) are shown for comparison. Frozen phonon total-energy as a function of the displacement of W atom for (c) α and (d) β phonon. Inset shows schematic illustration of the vibrational pattern in real space for α and β modes. Black dashed lines are harmonic fitting and blue solid lines are harmonic + quartic fitting. Both phonons are quite harmonic when the atom displacements are smaller than 0.1 Å.

TAB. 1. Comparison of DOS at Fermi level $N(E_F)$ (states/spin/Ry/unit cell), logarithmic average of the phonon frequencies $\omega_{\log}$(K), EPC constant $\lambda$ and superconducting transition temperature $T_c$ (K) of bulk WN, WN films, bulk W, WC and TaN. The $z$ in brackets denote the uniaxial pressure along $z$ axis. Experimental observed $T_c$ of bulk W, metal deposited WC and pressured WC are given for comparison.

| material | condition | $N(E_F)$ | $\omega_{\log}$ | $\lambda$ | $T_c$ |
|---|---|---|---|---|---|
| 3D WN | 0 GPa (non SOC) | 6.24 | 302.12 | 1.37 | 31.38 |
| | 0 GPa (with SOC) | 6.17 | 302.47 | 1.32 | 30.30 |
| | 43 GPa | 5.65 | 364.82 | 1.14 | 30.79 |
| WN film (2D $W_3N_4$) | pristine | 12.37 | 194.07 | 0.91 | 11.56 |
| WN film (2D $W_2N_3$) | pristine | 13.62 | 155.79 | 1.81 | 24.58 |
| 3D W | 0 GPa | 2.96 | 221.89 | 0.27 | 0.04 |
| | 0 GPa[49] | - | - | - | 0.015[exp] |
| 3D WC | pristine | 2.04 | 516.00 | 0.14 | 0.00 |
| | 0.2 e/unitcell | 2.00 | 342.41 | 0.11 | 0.00 |
| | 0.6 e/unitcell | 5.06 | 349.33 | 0.44 | 2.39 |
| | 1.0 e/unitcell | 5.74 | 342.61 | 0.54 | 5.57 |
| | metal deposition[20] | - | - | - | 3-12[exp] |
| | 35 GPa | 1.91 | 546.56 | 0.12 | 0.00 |
| | 28 GPa (z) | 2.19 | 525.06 | 0.14 | 0.00 |
| | 50 GPa (z) | 2.30 | 524.30 | 0.14 | 0.00 |
| 3D TaN | pristine | 0.52 | 619.08 | 0.02 | 0.00 |
| | 35 GPa | 0.54 | 690.22 | 0.03 | 0.00 |
| | 35 GPa [18] | - | - | - | ~4[exp] |
| | 0.2 e/unitcell | 3.74 | 319.24 | 0.41 | 1.48 |
| | 0.6 e/unitcell | 7.33 | 197.77 | 1.86 | 26.98 |
| | 1.0 e/unitcell | - | - | - | unstable |
| | 28 GPa (z) | 0.70 | 598.63 | 0.04 | 0.00 |
| | 50 GPa (z) | 1.00 | 562.37 | 0.06 | 0.00 |

As one of the most important quantities in TSC [51], the superconducting gap is obtained by solving the fully anisotropic Migdal-Eliashberg equations self-consistently using EPW code. Fig. 2(a) explicitly displays that bulk WN is a full gap superconductor, fulfilling one of the requirement of TSC. The perfect fitting of wannier interpolated bands with DFT bands (Fig. S4 in Supplemental Material) guarantees the reliability of EPW calculations. Electron pockets centered around $A$ which are dominated by W $d_{xy}$ orbitals (Fig. S5 in Supplemental Material) have stronger EPC, this leads to formation of the larger gap ~9 meV. Electron pockets centered around $K$ which are dominated by W $d_{xz}/d_{yz}$ orbitals (Fig. S5 in Supplemental Material) have weaker EPC, this results in the smaller gap ~5 meV. The superconducting gap value of WN are slightly larger than that of famous two-gap conventional superconductor $MgB_2$ whose $\Delta(0)$=6.8 meV for σ sheets and $\Delta(0)$=1.8 meV for π sheets and a $T_c$ of 39 K [33]. According to the BCS theory, the relation between the gap and the critical temperature holds as $k_B T_c = 2\Delta_0/3.52$. This suggest that after considering anisotropical coupling $T_c$ of WN can be even higher than 39 K.

The EPC in WN is so abnormally strong that we have to put our predictions on a firm playground. We check the influence of SOC and anharmonicity on EPC. SOC change the EPC significantly in some materials but from different aspects. For instance, SOC in $CaBi_2$ change the topology and the areas of the Fermi surface, harden some phonon modes and reduce the nesting, leading to a 50% reduction of total EPC strength [52]. Differently, SOC in bulk Pb has little effect on the electronic structure and Fermi surface, but induce obvious softening of all phonons, total λ is increased by 44% [53]. However, the DOS of WN at $E_F$ is reduced by only 1% when including SOC. From the calculated band structures and Fermi surfaces with SOC (Fig. S6 in Supplemental Material), we can see that all bands along the high symmetry lines are split except the band along $A$-$L$ by SOC, SOC lift the degeneracy of bands with a huge splitting energy of 0.58 eV, but these states are far away from Fermi level (1.32 eV below the Fermi level), which are irrelevant with EPC. The overall topology and area of Fermi surfaces are the same with those without SOC (with only a slight splitting). The phonon spectra calculated with inclusion of SOC (blue solid lines) are also identical with those calculated without considering SOC (black dashed lines) as shown in Fig. 2(b). The total difference of λ is only 0.05. Therefore, total EPC are unaffected by SOC for WN.

Anharmonicity is another factor that should be treated carefully. For instance, anharmonicity strongly renormalize the EPC and suppress the total λ and $T_c$ of $H_2S$ by 30% at 200 GPa [54]. Anharmonicity also reduces the total λ by more than 30% and $T_c$ by more than 59% in the range of 109-151 GPa for $AlH_3$ as compared with harmonic approximations [55]. For bulk WN, its EPC are dominated by vibrations of tungsten atom (183.8 u), which is much heavier than Al (26.98 u), S (32.07 u) and H (1.01 u) mentioned above. Most importantly, WN are stable under ambient pressure, meaning that the displacement of atoms are small enough to lie in the harmonic approximations. To quantitatively investigate the effect of anharmonicity, frozen-phonon total energy as a function of the displacement of W atom for α and β phonon are provided in Fig. 2(c,d). For α phonon, the distortion energy can be well fitted by the harmonic expression $E_\delta = 7.0\ \delta^2$ eV/Å$^2$ up to a displacement of 0.15 Å. For β phonon, the distortion energy begin to deviate from harmonicity when the atom displacement is larger than 0.1 Å. The distortion energy can be well fitted by the harmonic + quartic expression $E_\delta = A_2\ \delta^2 + A_4\ \delta^4$ eV/Å$^2$ = 3.4 $\delta^2$ + 34 $\delta^4$ eV/Å$^2$. The ratio of $A_4/(A_2*A_2) = 2.94$ is much smaller that of $E_{2g}$ mode ($A_4/(A_2*A_2) = 8$) in $MgB_2$ [45]. Even in $MgB_2$ the anharmonic $E_{2g}$ mode reduce the overall λ only by 10% [43]. Anharmonicity is believed to be important in materials containing light elements under high pressures (i.e. with large atomic displacement). To summarize this part, the EPC in WN remains unchanged when considering SOC and anharmonicity.

## B. Stability and superconductivity in 2D monolayer $W_3N_4$

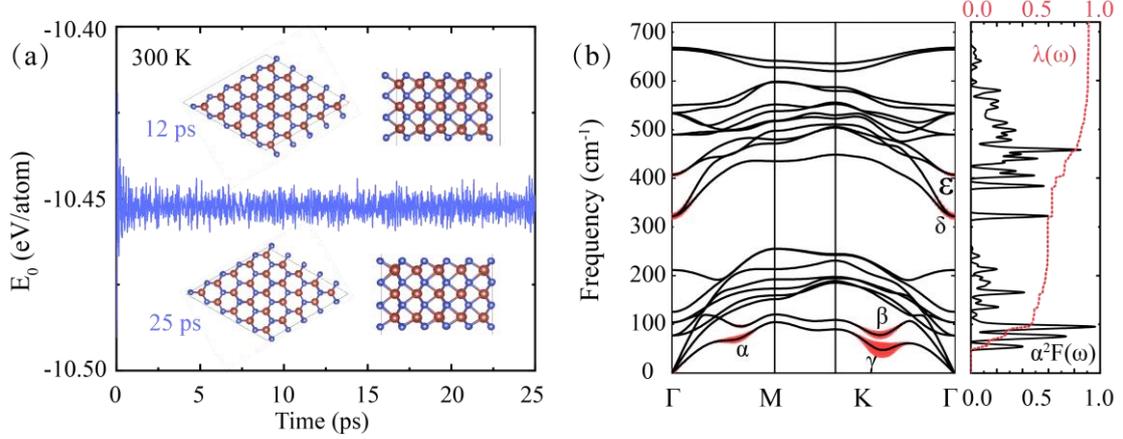

FIG. 3. (a) Variation of total potential energy $E_0$ with time during *ab initio* molecular simulation of monolayer $W_3N_4$ at 300 K, and top and side view of the structure at 12 ps and 25 ps during simulation. (b) Phonon dispersion decorated with red dots proportional to the partial EPC strength $\lambda_{\mathbf{q}\nu}$. (c) $\alpha^2F(\omega)$ (black lines) and cumulative frequency-dependent $\lambda(\omega)$ (red lines).

After the discussion of bulk materials, we next turn to the 2D limit of these compounds. First, the electronic structures of 2D films made from bulk WN with different thickness and terminations are presented in Fig. S7 in Supplemental Material. For Nitrogen terminated monolayers, the electronic bands changes dramatically from $WN_2$ to $W_4N_5$, then the bands of $W_5N_6$ are quite similar with $W_4N_5$. In contrast, the W terminated monolayers show strong resemblance from the thinnest $W_2N$ to the thickest $W_4N_5$ investigate in our work, the only difference is the relative energy of the band maximum at *M* and the band minimum at *K*. From these monolayers, we can obtain normal metals like $W_2N$ and $W_4N_3$, topological Dirac semimetal $W_2N_3$ with Dirac point 1.8 eV below the $E_F$ as demonstrated in ref [56] and topological nodal lines semimetal $W_3N_4$. However, emergence of topological nontrivial states are not universal. We are unable to find Dirac points or nodal lines near Fermi level in WC and TaN films with different thickness and terminated atoms, as shown in Fig. S8-9 in Supplemental Material. In this work, we focus on monolayer $W_3N_4$. A*b initio* molecular dynamic simulations are performed at 300 K using a 5 ×5 ×1 supercell to check the thermal stability, where the simulation time is set to 25 *ps* with a time step of 1.0 *fs*. As presents in Fig. 3(a), the evolution of total energy during the molecular dynamic simulations fluctuate slightly around a constant, the structures at 12 *ps* and 25 *ps* remain nearly the same atomic positions of the original structure, demonstrating high thermal stabilities at ambient condition and strong bonding in monolayer $W_3N_4$. The non-imaginary phonon dispersion in Fig. 3(b) demonstrate dynamical stability of monolayer $W_3N_4$. Furthermore, the elastic constants of monolayer $W_3N_4$ (see Tab. S2 in Supplemental Material) comply well with requirements of the mechanical stability criterion of a 2D materials [57]: $C_{11}*C_{22}-C_{12}*C_{21} > 0$ and $C_{66} > 0$, confirming its mechanical stability. In a word, monolayer $W_3N_4$ fulfill all the stability requirements.

Our previous work demonstrates that monolayer $W_2N_3$, which can be regard as the 2D monolayer composed by two layers of W and N from bulk WN, exhibit strong EPC and a predicted $T_c$ of 38 K [58]. This illuminate us to see whether the strong EPC is ubiquitous in WN film with different thickness. Fig. 3(b) presents the phonon dispersion decorated with red dots proportional to $\lambda_{\mathbf{q}\nu}$. (c) $\alpha^2F(\omega)$ (black lines) and cumulative frequency-dependent $\lambda(\omega)$ (red lines). The total $\lambda$ is as high as 0.91 and the predicted $T_c$ is 11.56 K. By inspecting the partial EPC strength $\lambda_{\mathbf{q}\nu}$ and cumulative frequency-dependent $\lambda(\omega)$, we are aware that the total EPC comes from two low lying acoustic modes which contribute 50% of the total $\lambda$ and the two optical modes from 318 cm$^{-1}$ to 458 cm$^{-1}$ which contribute 25% of the total $\lambda$. Compared with monolayer $W_2N_3$, the softening of the two acoustic mode in $W_3N_4$ are reduced obviously, therefore the $\lambda$ of these two modes is nearly half of that in monolayer $W_2N_3$. The optical modes above the frequency gap soften more remarkably at $\varGamma$ than monolayer $W_2N_3$. On should note that the distribution of the large $\lambda_{\mathbf{q}\nu}$ in Brillouin zone are quite similarly with that of monolayer $W_2N_3$ [58], because the topology of Fermi surfaces and bonding properties in monolayer $W_3N_4$ and $W_2N_3$ resembles. Those phonons exhibit the largest $\lambda_{\mathbf{q}\nu}$ are the most soft phonons as denoted by α, β, γ, δ and ε. α phonon involves our-of-plane vibration of all W atoms in the same directions, while the N atoms are nearly silent. For β (γ) phonon, only two outer W atoms move in the opposite (same) direction along out-of-plane directions. Higher frequency δ and ε phonons are associated with in-plane vibrations of N atoms in which δ phonons break, while ε phonons preserve the out-of-plane mirror symmetry (see Fig. S10 in Supplemental Material).

## C. Topological nodal lines in 2D $W_3N_4$

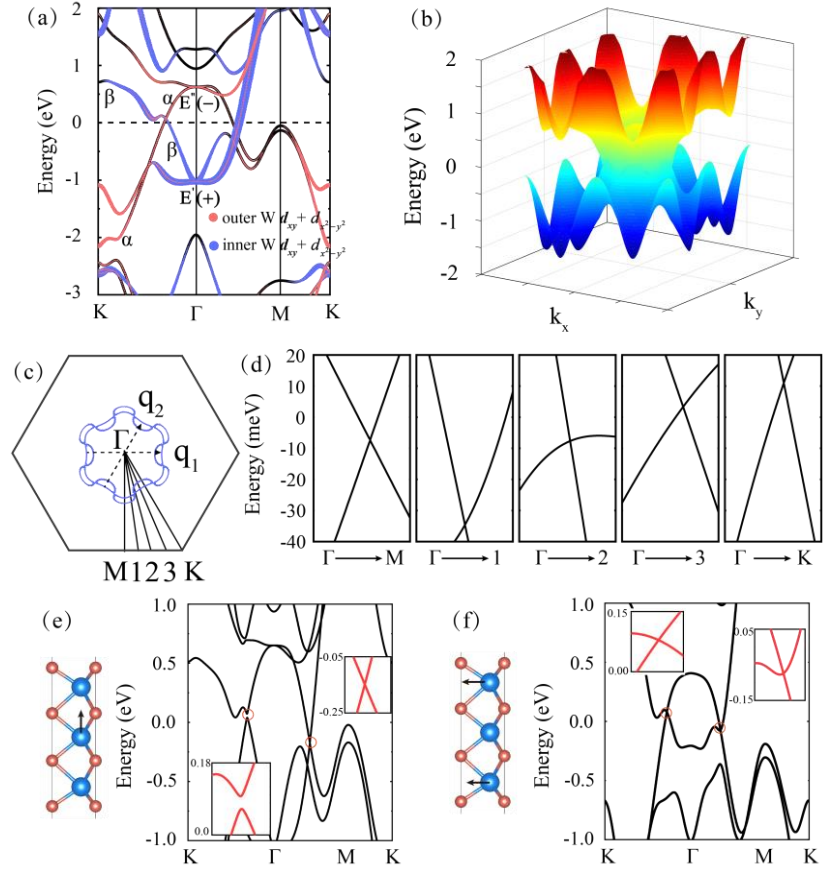

FIG. 4. (a) Orbital projected band structures of monolayer $W_3N_4$, the crossing bands forming nodal line are labeled as α and β. The representation of α and β at $\Gamma$ point are $E''(-)$ and $E'(+)$ respectively, in which the "−" and "+" denote the mirror eigenvalue of the bands. The weights of inner W $d_{xy}+d_{x^2-y^2}$ (outer W $d_{xy}+d_{x^2-y^2}$) states are proportional to the size of the blue (red) dots. (b) Three dimensional (3D) band structure near Fermi level showing the crossing of α and β bands in the whole Brillouin zone. (c) Fermi surface of monolayer $W_3N_4$. To explicitly show nodal ring the plotting is set with energy window [−20 meV, 10 meV]. (d) DFT calculated band structures along the lines shown in (c) with dense $k$ points. (e) Band structures of monolayer $W_3N_4$ with the mirror symmetry broken by shifting the inner W atom slightly away from the equilibrium position inner W move along $z$, inset shows the gap opening along $\Gamma$-$K$ and gapless crossing along $\Gamma$-$M$. (f) Band structures of monolayer $W_3N_4$ with only the mirror symmetry preserves by shifting the two outer W atom away from the equilibrium position along $x$ direction with the same displacement. inset show gapless crossing along $\Gamma$-$K$ and $\Gamma$-$M$. All atom displacements are 0.13 Å.

In this section, we focus on the electronic structure of monolayer $W_3N_4$. It is noncentrosymmetric as a whole but reflection symmetric with respect to the inner W plane. As shown in Fig. 4(a), the bands around the Fermi level mainly arise from inner W $d_{xy}+d_{x^2-y^2}$ and outer W $d_{xy}+d_{x^2-y^2}$ states. The inner W $d_{xy}+d_{x^2-y^2}$ bands are higher in energy than the outer W $d_{xy}+d_{x^2-y^2}$ states, but a band inversion happens around the $\Gamma$ point. Along $K$-$\Gamma$ and $\Gamma$-$M$ lines, α and β bands linearly cross with each other at two points and form two Dirac points near the Fermi level. The we obtain the representation of bands by Irvsp package [59], The little group of $\Gamma$ point is $D_{3h}$, α band belongs to $E''$ representation with twofold degeneracy and mirror eigenvalue of -1. β band belongs to $E'$ representation with twofold degeneracy and mirror eigenvalue of +1. At the points between $\Gamma$-$M$, the little group is $C_{2v}$, α band belongs to $B_2$ representation with no degeneracy and mirror eigenvalue of -1, β band belongs to $B_1$ representation with no degeneracy and mirror eigenvalue of +1. At the points between $\Gamma$-$K$ (and at $K$), the little group of is $Cs(C_{3h})$, α band belongs $A''$ ($E''$) representation with no degeneracy and mirror eigenvalue of -1, β band belongs to $A'$ ($E'^*$) representation with no degeneracy and mirror eigenvalue of +1. We also check other points in the Brilloun zone, results show that the α and β present opposite mirror parity at any points. Therefore, the crossing are protected by mirror symmetry. According to the widely used [60,61] classification of nodal lines proposed by *C. Fang* [62], the nodal lines can be characterized by a topological quantum number $\zeta_0=N_1-N_2$, where $N_1$ and $N_2$ are the number of bands below the Fermi energy that has mirror eigenvalue of +1 at the two points $p_1$ and $p_2$ inside and outside the nodal line. In our case, $p_1$ and $p_2$ are chosen as $\Gamma$ and $K$ respectively. $N_\Gamma$=34, $N_K$=33, finally we get $\zeta_0$=+1, confirming that the nodal lines are topologically nontrivial. In Fig. 4(b), 3D band structures show that the VB and CB linearly cross in all directions around the Fermi level and thus result in a closed ring around the $\Gamma$ point. To visualize the nodal rings, we plot the Fermi surface of monolayer $W_3N_4$ including the bands with the energy range from $E_F$-20 meV to $E_F$+10 meV as shown Fig. 4(c). We check the band crossing along $\Gamma$-$M$, $\Gamma$-$1$, $\Gamma$-$2$, $\Gamma$-$3$ and $\Gamma$-$K$ by calculating band energies with very dense grid (see Fig. 4(d)). The location of the crossing points are the highest ($E_F$+10 meV) along $\Gamma$-$K$ and lowest ($E_F$-30 meV) near $\Gamma$-$1$, the energy width of ~40 meV is narrower than those of InNbS$_2$ (80 meV), InNbSe$_2$ (780 meV) [63], Cu$_3$PdN (200 meV) [64] and bulk beryllium (800 meV) [65], indicating the nodal lines are quite flat. To examine the role of symmetry in protecting the nodal line, we calculate bands of artificially distorted structures. The distortion in Fig. 4(e) reduce the symmetry to *P3m1* (*No*. 156) space group, horizontal mirror symmetry is broken, the crossing between *Γ-K* is gapped with gap value of 32 meV, the crossing along *Γ-M* are still gapless. By shifting the two outer W atom away from the equilibrium position along *x* direction with the same displacement. The distortion in Fig. 4(f) reduce the symmetry to *Pm* (*No*. 6) space group, the only symmetry operations exist are identity and horizontal mirror symmetry. The little group of *Γ* point is *C$_s$*, the two bands forming the crossing belong to representations of *A$'$* and *A$''$* with opposite mirror eigenvalues, thus the nodal lines are intact in this structure. This testing undoubtedly confirm that horizontal mirror symmetry is sufficient to protect the nodal lines.

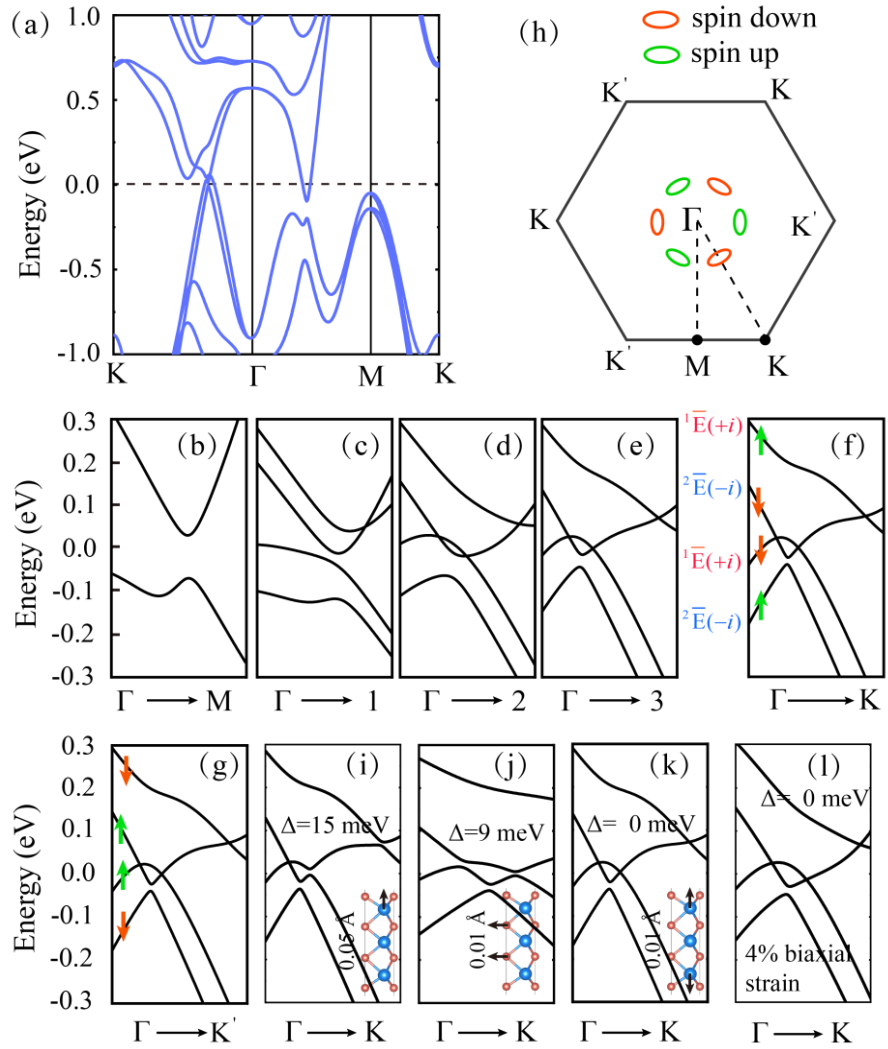

FIG. 5. (a) Band structure of monolayer W$_3$N$_4$ calculated with the inclusion of SOC. Blowup of the band structure near the crossing along (b) *Γ-M* (c) *Γ-1* (d) *Γ-2* (e) *Γ-3* (f) *Γ- K* and (g) *Γ- K'*. The representation of the four bands in (f) are shown on the left, with +*i* and –*i* as its corresponding mirror eigenvalues. The red and green arrows in (f) and (g) denotes the spin down and up orientations along *z* axis. (h) Schematic picture of the nodal lines in monolayer W$_3$N$_4$ when including SOC. Blowup of the band structure near the crossing (i) when one outer

W atom is shifted along *z* direction, (j) two inner N atoms are shifted along *x* direction, (h) two outer W atoms are shifted oppositely with the same distance to preserve the original crystal symmetry, (l) applying 4% tensile strain.

On one hand, SOC always exists in real materials, on the other hand, electronic structures with SOC included are indispensable to classify their topological nature. For mirror reflection symmetry protected nodal lines, depending on the atomic orbitals making up the electronic states and the strength of SOC, SOC can turn a spinless Dirac nodal lines into a spinful Weyl nodal rings within the mirror planes (i.e. $InNbS_2$ [63]); or a Weyl semimetal which has Weyl points off the mirror planes (i.e. TaAs [66] and $InNbSe_2$ [63]); or a topological insulator when the strength of SOC is strong enough (i.e. $Hg_3As_2$ [67], HfC [68]). There is a large gap opening along *Γ-M* in monolayer $W_3N_4$ (see Fig. 5(a)). To check the crossing in the whole Brillouin zone, we recalculate the bands along *Γ-M*, *Γ-1*, *Γ-2*, *Γ-3*, *Γ-K* and *Γ-K′* as show in Fig. 5(b-g) with dense **k** points. The crossing are gapless along *Γ-2*, *Γ-3*, *Γ-K* and *Γ-K′*. The component of spins of the two crossing bands are mainly out-of-plane directions. In addition, the spin direction along *Γ-K* and *Γ-K′* are opposite as denoted by the arrows in Fig. 5(f,g). This may lead to spin-orbit locking induced Ising superconductivity, which can sustain large in-plane magnetic field [69]. Due to the existence of the horizontal mirror symmetry, we can classify these four bands by mirror eigenvalues of $\pm i$ which are illustrated in Fig. 5(f,g). At the gapless crossing position, the two bands have opposite mirror eigenvalues and thus the band crossing is protected by the mirror symmetry. We schematically display the nodal lines in Fig. 5(h) when SOC is included with red and green denote spin-down and spin-up rings. So far, only a limited number of materials are proposed as nodal lines semimetals that are robust against SOC [25-27], therefore the nodal lines found here is precious. In addition, the nodal lines in monolayer $W_3N_4$ are clean (namely, without mixing of other nontrivial states), this makes the exotic topological properties easily accessible in experiment. To check the survival of these spinful nodal lines. We move one W atom along *z* direction (the reflection symmetry is broken), the crossing is gapped as shown in Fig. 5(i), indicating that reflection symmetry is necessary. Next, we distort the structure when there is only reflection symmetry (Fig. 5(j)), the nodal line are also gapped, demonstrating that reflection symmetry its self can not ensure the crossing which is different from the case of non-SOC. Keeping the group symmetry of pristine structure and move the outer W atom along *z* direction oppositely (Fig. 5(k)), or even if applying 4% biaxial tensile strain (Fig. 5(l)), the crossing remains gapless. Above two tests declare that the crossing are not sensitive to orbital interaction strength (but it matters).

## IV. CONCLUSION

In summary, through exhaustive first-principles calculations, we discover that the three-component fermion candidate WN exhibit overwhelmingly strong EPC and can be a superconductor with relatively high transition temperature of 31 K. Through comparison of Fermi surface, real space charge distribution, phonon dispersion and EPC between WN, WC, TaN, MoP and elemental metal tungsten, we unravel that the unique and extremely strong EPC in WN results from strong Fermi nesting and large deformation potential. In contrast, pristine WC and TaN has a negligible small EPC. Electron doping can efficiently increase the EPC strength of WC and TaN and the predicted $T_c$ are in the same order of experiments, suggesting that the superconductivity found in experiments may be explained by electron-doping enhanced EPC. Based on our results, we propose that the Fermi nesting and partial charge distribution at Fermi level, both of which are computationally cheap, can be used as two preliminary criteria for high-throughput screening of novel superconductors.

Go one step further, we find that the strong EPC and superconductivity are ubiquitous in WN films. Most importantly, monolayer $W_3N_4$ uniquely host topological nodal lines that are flat in energy and momentum space and lies closely near the Fermi level In addition, there are no other nontrivial bands near Fermi level and these nodal lines are robust against spin orbit coupling, which is rare in real materials and making the exotic topological properties easily accessible. Finally, the coexistence of superconductivity (with high transition temperature and full gap) and topological states in WN make it a promising materials for realizing topological superconductivity. As the 2D film of WC and WN are already obtained by experiments, monolayer $W_3N_4$ provide a good platform for exploring 2D superconductivity and their interplay with topological states.

**Data availability**

All relevant data are available from the corresponding author

## ACKNOWLEDGMENTS

This work was supported by the National Natural Science Foundation of China (Grant No. 51272291)，Guangxi Natural Science Foundation (Grant No. 2019GXNSFBA245077) and the Scientific Research Fund of Guilin University of Aerospace Technology. J. Chen acknowledges the support from Thousand Teacher Programm of Guangxi. The authors thank Wenzhe Zhou and Fangping Ouyang in Central South University for supports and helpful discussions.

**AUTHOR CONTRIBUTIONS**

Jianyong Chen designed the project. Jiacheng Gao calculate and analyze the band representations. All authors wrote and revised the manuscript.

**Competing interests:** The Authors declare no Competing Financial or Non-Financial Interests.